%% file: hadron2011.tex
\documentclass[a4paper,11pt]{article}

\usepackage{contribution}

\input{econfmacros}
\input{contributionmacros}

\begin{document}

\input{contribution}

\end{document}

%% file: econfmacros.tex
\newcommand{\weblink}[2][]{%
    \ifthenelse{\equal{#1}{}}%
    {\textnormal{\url{#2}}}%
    {\textnormal{\href{#2}{#1}}}%
}

\newcommand{\acknowledgements}[1]{%
  \bigskip\bigskip
  \textsf{\textbf{\Large Acknowledgements}} \\[2ex]
  {#1}
  \bigskip
}

\def\beq{\begin{equation}}
\def\eeq#1{\label{#1}\end{equation}}
\def\eeqn{\end{equation}}

\def\beqa{\begin{eqnarray}}
\def\eeqa#1{\label{#1}\end{eqnarray}}
\def\eeqan{\end{eqnarray}}

\let\bar=\overbar

\def\Dslash{\not{\hbox{\kern-4pt $D$}}}
\def\dslash{\not{\hbox{\kern-2pt $\del$}}}

\def\msb{{\bar{\ssstyle M \kern -1pt S}}}

%% file: contributionmacros.tex
\newcommand{\contribution}[7][]{%
  \clearpage
  \thispagestyle{plain}
  \ifthenelse{\equal{#1}{}}
  {\hypersetup{pdftitle={#2}}}
  {\hypersetup{pdftitle={#1}}}
  \hypersetup{pdfauthor={{#3} {#4}}}
  {\centering\normalfont\LARGE\bfseries\sffamily #2 \par\nobreak}
  \lhead{}
  \chead{%
    \textit{\footnotesize XIV International Conference on Hadron Spectroscopy
      (\weblink[\textit{hadron2011}]{http://www.hadron2011.de}), 13-17 June 2011, Munich, Germany}%
  }
  \rhead{}
  \bigskip
  \begin{center}
    {#3} {#4}\ifthenelse{\equal{#6}{}}{}{\footnote{\weblink[#6]{mailto:#6}}}
    \ifthenelse{\equal{#7}{}}{}{#7} \\
    \textit{#5}
  \end{center}
  \bigskip
}

\renewcommand{\abstract}[1]{%
  \begin{center}
    \begin{minipage}{0.85\textwidth}
      \begin{footnotesize}
        #1
      \end{footnotesize}
    \end{minipage}
  \end{center}
  \bigskip
}

%% file: contribution.tex
%
%
%
%
%
{  


%

\contribution{Mass dependence of the heavy quark potential\\
  and its effects on quarkonium states}  
{Alexander}{Laschka}  
{Physik Department\\
  Technische Universit\"{a}t M\"{u}nchen\\
  D-85747 Garching, GERMANY}  
{}  
{\!\!, Norbert Kaiser, and Wolfram Weise}
%

\abstract{%
  The heavy quark-antiquark potential is accessible in perturbative QCD and in
  lattice simulations. The perturbative short-distance part of the potential is
  contructed via a restricted Fourier transform, covering the momentum region
  where perturbative QCD is applicable. We show that for the leading order
  static term as well as for the mass dependent corrections, the perturbative
  part can be matched at intermediate distances with results from lattice QCD.
  From these matched potentials, quarkonium spectra with a single free parameter
  (the heavy quark mass) are derived and compared with empirical spectra.
  Furthermore, charm and bottom quark masses are deduced.
}
%

\section{The static potential}

The potential between two heavy quarks is a prime subject of interest since the
early days of QCD. Nowadays it is defined in a non-relativistic effective theory
framework. While the long distance part can be studied in lattice QCD
simulations, perturbation theory should be expected to work at short distances.
The potential can be organized in a power series of the inverse quark mass $m$:
\begin{equation}
  \label{eq:potential_expansion}
  V=V^{(0)}+\frac{V^{(1)}}{m/2}+\frac{V^{(2)}}{(m/2)^2}+\ldots\, .
\end{equation}
The leading term $V^{(0)}$ represents the static potential. It has the following
form at two-loop order in momentum space:
\begin{equation}
  \label{eq:static_potential_1}
  \tilde{V}^{(0)}(|\vec q\,|) =-\frac{16\pi \alpha_s(|\vec q\,|)}{3\vec q\,^2}\,
  \bigg\{1+\frac{\alpha_s(|\vec q\,|)}{4\pi}\, a_1
  +\left(\frac{\alpha_s(|\vec q\,|)}{4 \pi}\right)^2 a_2
  +\ldots \bigg\}\, ,
\end{equation}
where $\vec q$ is the three-momentum transfer. The constants $a_1$ and $a_2$
are~\cite{Peter:1996ig,Peter:1997me,Schroder:1998vy}:
\begin{align}
  a_1 &= 31/3 - 10/9\,n_f,\\
  a_2 &= 456.749-66.3542\,n_f+1.23457\,n_f^2,
\end{align}
where $n_f$ is the number of light-quark flavors. Higher order terms have
infrared contributions and are not considered at this level. Expressing
$\alpha_s(|\vec q\,|)$ in a power series expansion about $\alpha_s$ at a fixed
scale $\mu$ leads to the standard definition of the $r$-space static potential,
\begin{multline}
  \label{eq:static_potential_2}
  V^{(0)}(r) = -\frac{4}{3} \frac{\alpha_s(\mu)}{r}\, \bigg\{1 
  + \frac{\alpha_s(\mu)}{4\pi}\,\Big[ a_1 + 2\beta_0\, g_\mu(r) \Big]\\
  + \bigg(\frac{\alpha_s(\mu)}{4\pi}\bigg)^2 \Big[ a_2 
  + \beta_0^2\left(4g_\mu^2(r)\! +\! \pi^2/3\right)
  + 2g_\mu(r)(2a_1\beta_0\!+\!\beta_1)\Big]
  +\mathcal O(\alpha_s^3) \bigg\} \, ,
\end{multline}
with $g_\mu(r)=\ln(\mu r)\!+\! \gamma_{\scriptscriptstyle\text{E}}\,$.
It is well known that this potential suffers from renormalon 
ambiguities~\cite{Beneke:1998rk,Hoang:1998nz} and shows a badly convergent
behavior~\cite{Pineda:2002se}.

We work in the following in the potential subtracted (PS) scheme proposed by
Beneke~\cite{Beneke:1998rk} and define the static $r$-space potential,
\begin{equation}
  \label{eq:r_space_potential}
  V^{(0)}(r,\mu_f) = \intop_{|\vec q\,|>\mu_f}\!
  \frac{d^3q}{(2\pi )^3}\ e^{i\vec q\cdot\vec r}\,
  \tilde{V}^{(0)}(|\vec q\,|)\, ,
\end{equation}
where $\tilde{V}^{(0)}(|\vec q\,|)$ is given in
Eq.~(\ref{eq:static_potential_1}), but $\alpha_s(|\vec q\,|)$ is understood
without resorting to a power series expansion. The momentum space cutoff $\mu_f$
is introduced in order to exclude the uncontrolled low momentum region.
\begin{figure}[tb]
  \begin{center}
  \includegraphics[width=.48\textwidth]{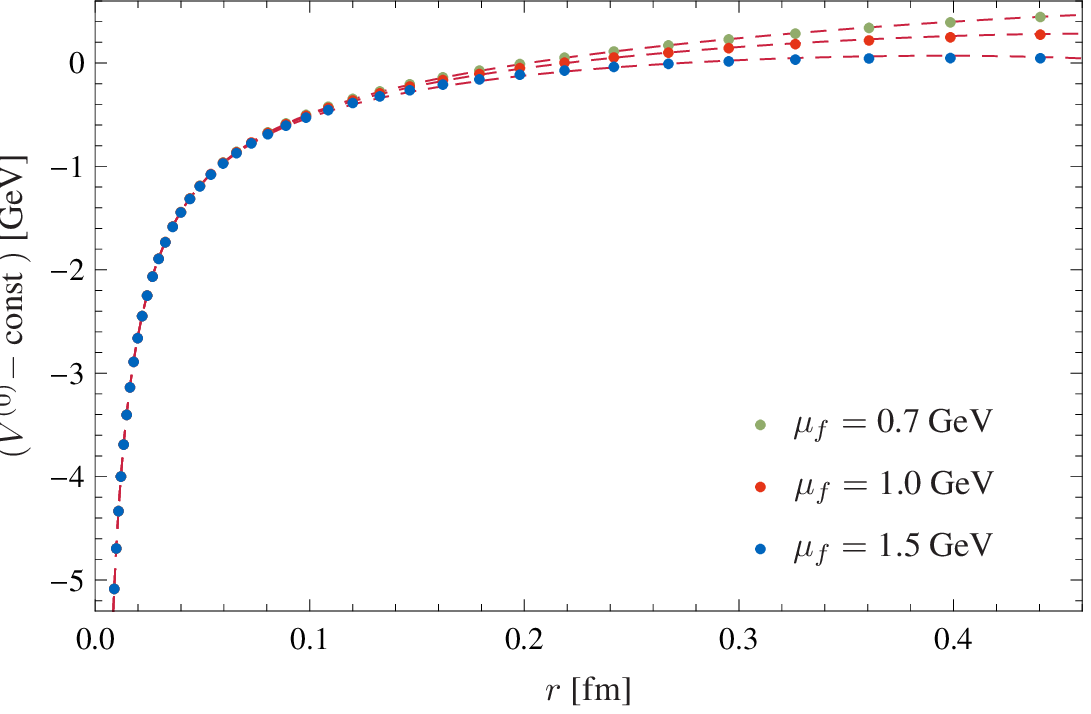}
  \hspace{1em}
  \includegraphics[width=.48\textwidth]{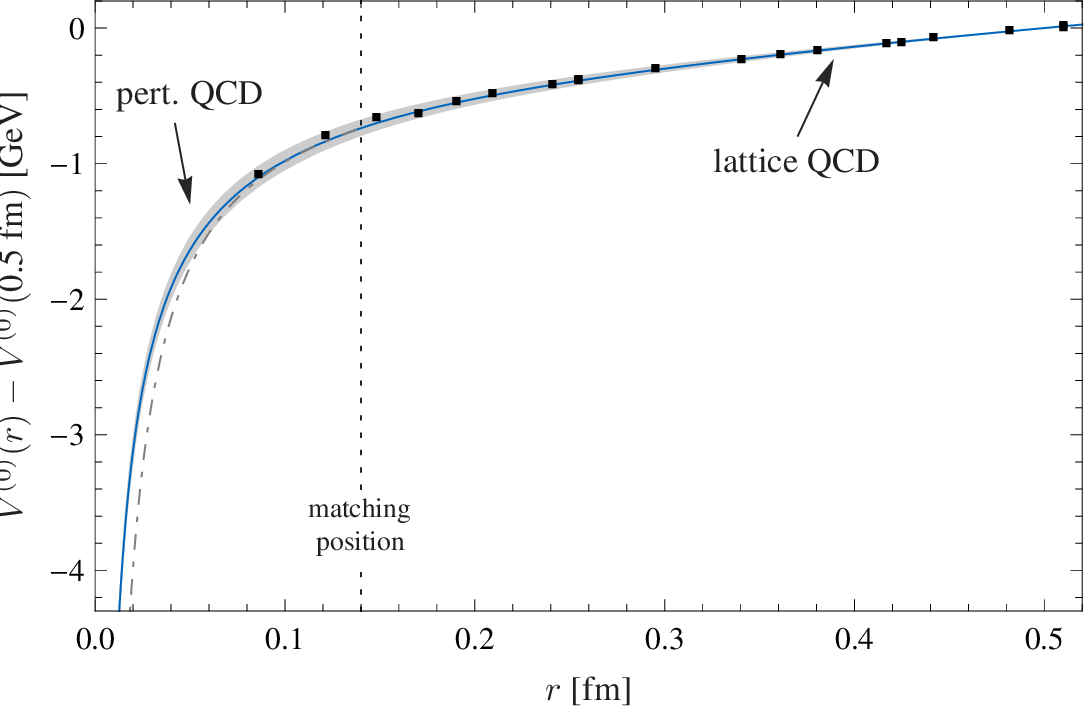}
  \caption{Static QCD potential (with $n_f=3$) from the restricted
  numerical Fourier transform~(\ref{eq:r_space_potential}). Left: coordinate
  space potential at two-loop order for different values of $\mu_f$. The curves
  have been shifted by a constant to match at small $r$ values. Right: potential
  matched at $r=0.14$~fm to a potential from lattice QCD~\cite{Bali:2000vr}.
  Taken from Ref.~\cite{Laschka:2011zr}.}
  \label{fig:1}
  \end{center}
\end{figure}
The potential $V^{(0)}(r,\mu_f)$ is evaluated numerically using four-loop RGE
running for the strong coupling $\alpha_s$. For distances $r\lesssim 0.2$~fm,
this potential depends only marginally on $\mu_f$ as shown in the left plot of
Fig.~\ref{fig:1}. The perturbative potential, valid at small distances, can be
matched at intermediate distances to results from lattice QCD (see the rightmost
plot in Fig.~\ref{fig:1}). For the matching point (dashed line) we choose
$r=0.14$~fm where both the perturbative and lattice potential are expected to be
reliable.

\section{The order $\mathbf{1/m}$ potential and quarkonium spectroscopy}

$V^{(1)}$ in Eq.~(\ref{eq:potential_expansion}) is the first mass dependent
correction to the static potential. It is spin independent and the leading term
reads in momentum space~\cite{Brambilla:2000gk}:
\begin{equation}
  \tilde{V}^{(1)}(|\vec q\,|) = -\frac{2 \pi^2 
  \alpha_s^2(|\vec q\,|)}{|\vec q\,|}\big\{1
  +\mathcal O(\alpha_s) \big\}\, .
\end{equation}
It can be transformed analogously as in Eq.~(\ref{eq:r_space_potential}) to
$r$-space with a low momentum cutoff $\mu'_f$. The dependence of $V^{(1)}$ on the
cutoff scale is again very weak for distances $r\lesssim 0.2$~fm. At long
distances $V^{(1)}(r)$ is known from lattice
QCD~\cite{Koma:2006si,Koma:confinement8}. To fit the lattice data we use the
form
\begin{equation}
  V^{(1)}_{\text{fit}}(r)=-\frac{c'}{r^2}+d' \ln\Big(\frac{r}{r_0}\Big)
  + \text{const},
\end{equation}
motivated in~\cite{PerezNadal:2008vm}.
As shown in Fig.~\ref{fig:2} matching with the perturbative potential at
intermediate distances is also possible at order $1/m$.
\begin{figure}[tb]
  \begin{center}
  \includegraphics[width=.48\textwidth]{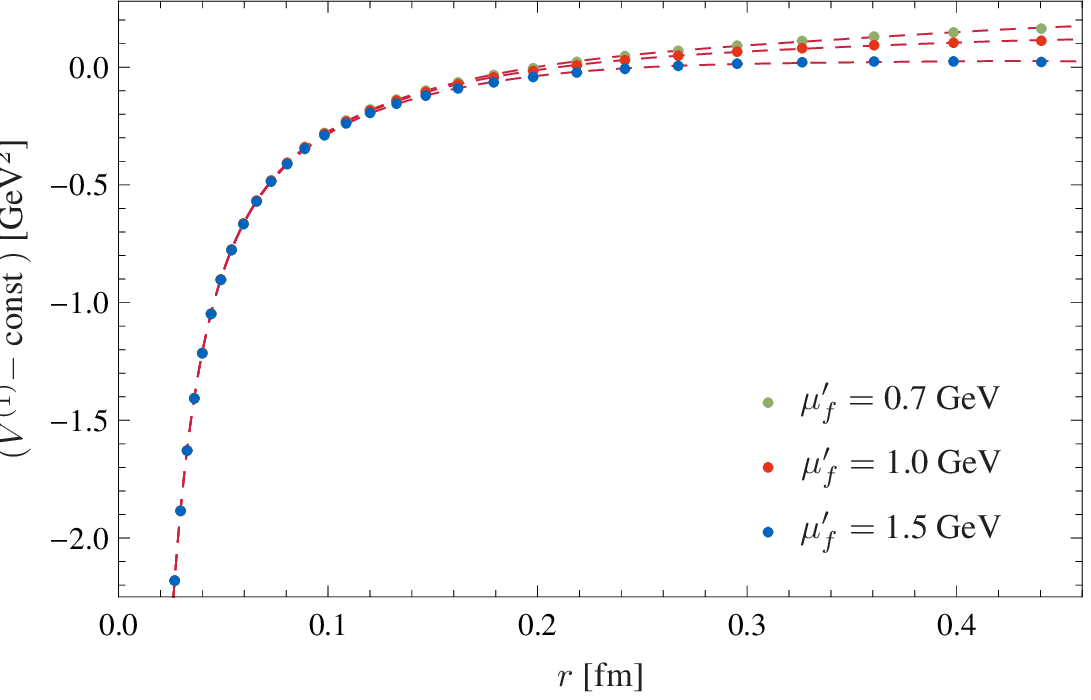}
  \hspace{1em}
  \includegraphics[width=.48\textwidth]{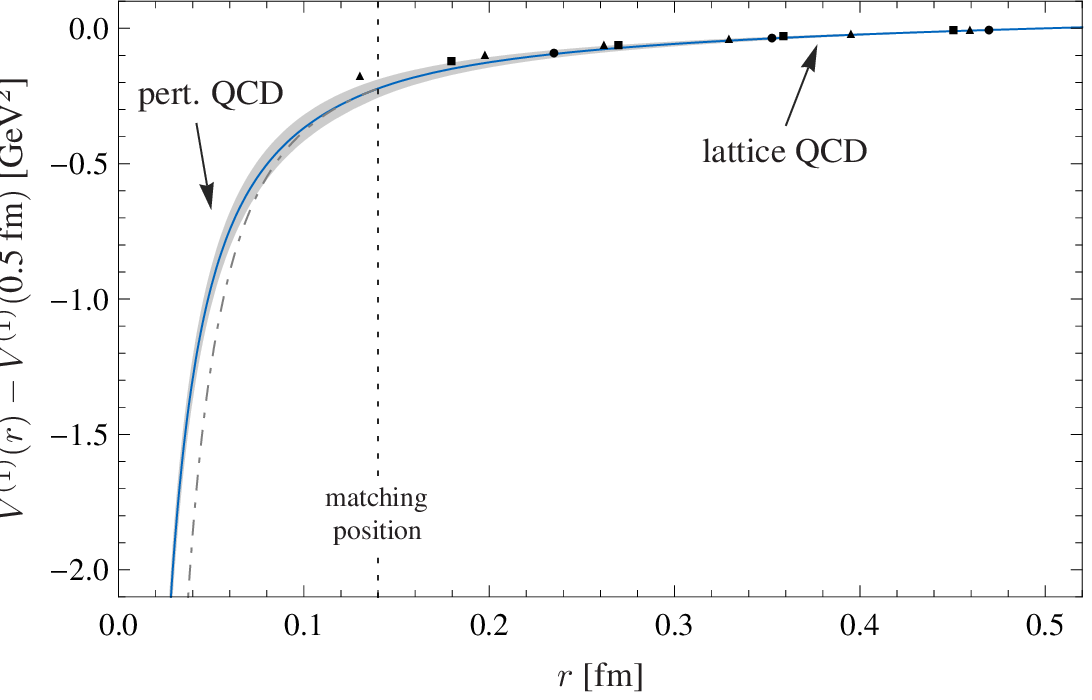}
  \caption{The order $1/m$ potential with $n_f=3$, from the restricted Fourier
  transform. Left: perturbative potential for different cutoffs $\mu'_f$.
  Right: perturbative potential matched at intermediate distances to a potential
  from lattice QCD. Taken from Ref.~\cite{Laschka:2011zr}.}
  \label{fig:2}
  \end{center}
\end{figure}

Using $V^{(0)}$ and $V^{(1)}$ as input in the Schr\"{o}dinger equation, we can
examine bottomonium and charmonium spectra. The overall constant of the
potential is the only free parameter. This single parameter is related to the
heavy quark mass in the PS scheme and can be translated in a second step to the
bottom and charm quark masses in the $\overline{\text{MS}}$ scheme
(see~\cite{Laschka:2011zr} for details). Our findings for the masses are
summarized in Table~\ref{tab:t1} and compared to the values listed by the
Particle Data Group (PDG)~\cite{Nakamura:2010zzi}.
\begin{table}[hbt]
  \begin{center}
  \begin{tabular}{lcc|c}
  \multicolumn{4}{c}{ $\overline{\text{MS}}$ masses [GeV]}\\
  \hline
  &Static&Static + $\mathcal O (1/m)$&PDG 2010\\
  \hline
  Bottom quark&$4.20\pm 0.04$
    &$4.18^{+0.05}_{-0.04}$
    &$4.19^{+0.18}_{-0.06}\vphantom{\Big(}$\\
  Charm quark&$1.23\pm 0.04$
    &$1.28^{+0.07}_{-0.06}$
    &$1.27^{+0.07}_{-0.09}\vphantom{\Big(}$\\
  \hline
  \end{tabular}
  \caption{Comparison of quark masses obtained in our approach (leading order
    plus order~$1/m$ corrections) with the values listed by the Particle Data
    Group (PDG)~\cite{Nakamura:2010zzi}.}
  \label{tab:t1}
  \end{center}
\end{table}
\begin{figure}[tb]
  \begin{center}
  \begin{minipage}{.465\textwidth}
    \begin{center}
    Bottomonium
    \end{center}
    \vspace{-1.5ex}
    \includegraphics[width=\textwidth]{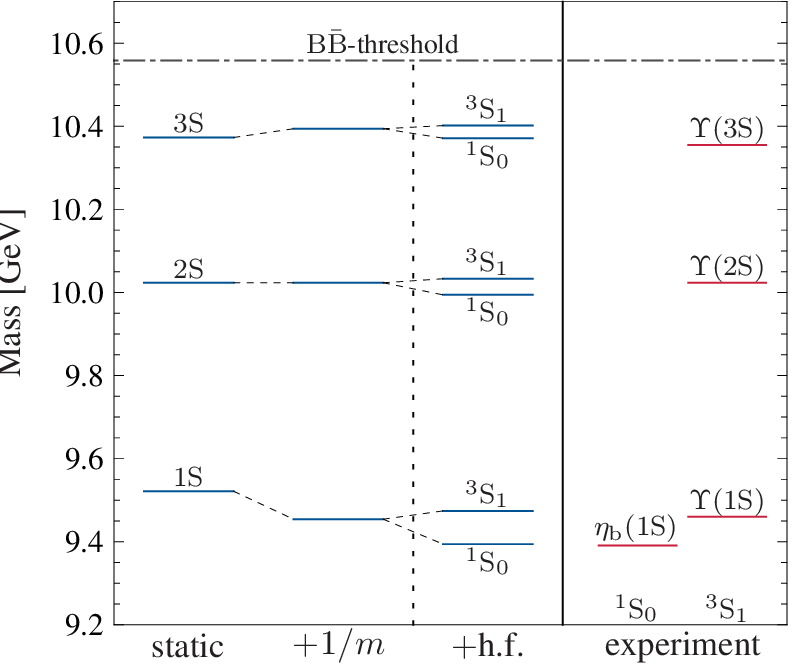}\\[0.5ex]
    \includegraphics[width=\textwidth]{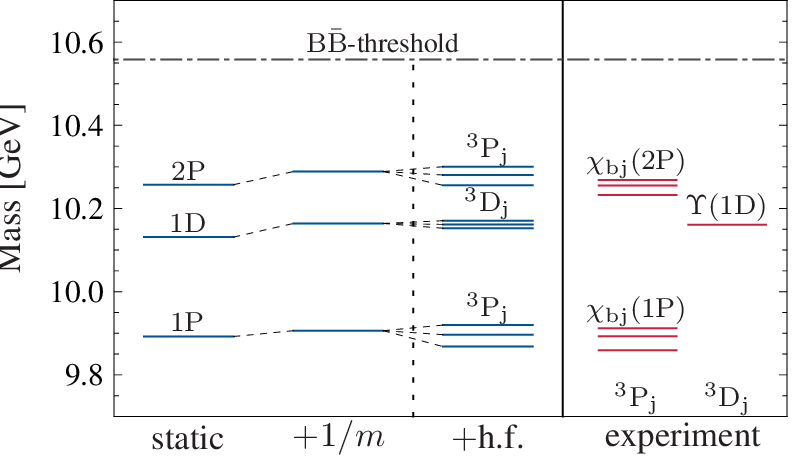}
  \end{minipage}
  \hspace{1.5em}
  \begin{minipage}{.484\textwidth}
    \begin{center}
    Charmonium
    \end{center}
    \vspace{-1.5ex}
    \includegraphics[width=\textwidth]{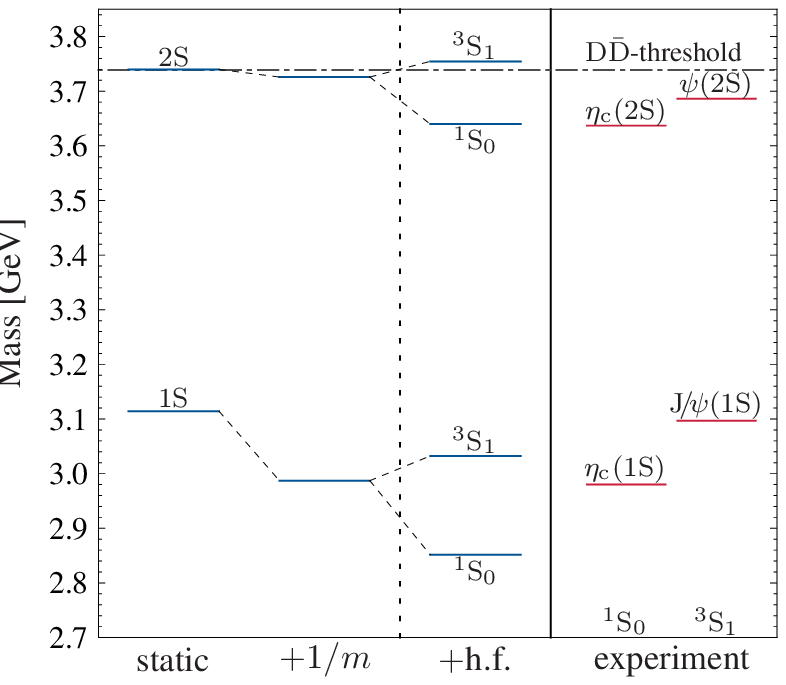}\\[0.5ex]
    \includegraphics[width=\textwidth]{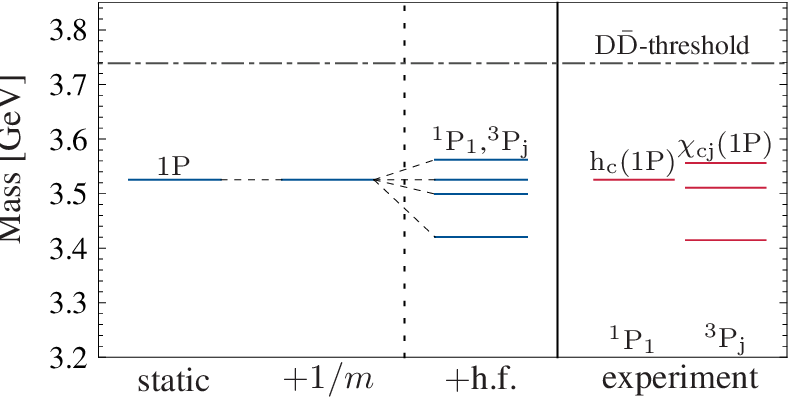}
  \end{minipage}
  \caption{Bottomonium and charmonium spectrum in comparison with experiment.
  Static plus order $1/m$ results are shown, with additional hyperfine effects
  (h.f.) added phenomenologically. Taken from Ref.~\cite{Laschka:2011zr}.}
  \label{fig:3}
  \end{center}
\end{figure}%
Results for the bottomonium and charmonium spectra are shown in
Fig.~\ref{fig:3}. In both cases we find that the 1S states are the most strongly
affected by $1/m$-effects. An additional effective one-gluon exchange spin
dependent term with $\alpha_s^{\text{eff}}=0.3$ (+h.f. in Fig.~\ref{fig:3}) would
improve our predictions. Of course, this step is purely ad hoc and needs to be
substituted by the full potential of order $1/m^2$, to be investigated in
forthcoming work.

\acknowledgements{%
  Work supported in part by BMBF, GSI and the DFG Excellence Cluster
  ``Origin and Structure of the Universe''.
}


%

}  
